\documentclass{article}
\usepackage{ijcai19}

\usepackage{times}
\usepackage{soul}
\usepackage{url}
\usepackage[hidelinks]{hyperref}
\usepackage[utf8]{inputenc}
\usepackage[small]{caption}
\usepackage{graphicx}
\usepackage{amsmath}
\usepackage{amsfonts}
\usepackage{booktabs}
\usepackage{makecell}
\usepackage{slashbox}
\usepackage[ruled,linesnumbered]{algorithm2e}
\usepackage{caption}
\usepackage{subcaption}
\usepackage{xcolor}
\usepackage{breqn}
\usepackage{scrextend}
\usepackage{enumitem}

\def\BibTeX{{\rm B\kern-.05em{\sc i\kern-.025em b}\kern-.08emT\kern-.1667em\lower.7ex\hbox{E}\kern-.125emX}}

\title{Deep Adversarial Network Alignment}

\author{
Tyler Derr$^1$
\and
Hamid Karimi$^1$\and
Xiaorui Liu$^{1}$\and
Jiejun Xu$^2$ \and
Jiliang Tang$^1$
\affiliations
$^1$\ Data Science and Engineering Lab, Michigan State University\
$^2$HRL Laboratories
\emails
\{derrtyle, karimiha, xiaorui\}@msu.edu,
jxu@hrl.com,
tangjili@msu.edu
}

\begin{document}

\maketitle

\begin{abstract}Network alignment, in general, seeks to discover the hidden underlying correspondence between nodes across two (or more) networks when given their network structure. However, most existing network alignment methods have added assumptions of additional constraints to guide the alignment, such as having a set of seed node-node correspondences across the networks or the existence of side-information. Instead, we seek to develop a general network alignment algorithm that makes no additional assumptions. Recently, network embedding has proven effective in many network analysis tasks, but embeddings of different networks are not aligned. Thus, we present our Deep Adversarial Network Alignment (DANA) framework that first uses deep adversarial learning to discover complex mappings for aligning the embedding distributions of the two networks. Then, using our learned mapping functions, DANA performs an efficient nearest neighbor node alignment. We perform experiments on real world datasets to show the effectiveness of our framework for first aligning the graph embedding distributions and then discovering node alignments that outperform existing methods.
\end{abstract}

\section{Introduction}

In today's world, networks are arising almost everywhere from social to biological networks. This has caused an increased attention in the domain of network analysis. However, most efforts have primarily focused on single network problems such as link prediction~\cite{liben-nowell2007linkpred} and community detection~\cite{fortunato2010commdetect}, but many problems inherently are only defined when having multiple networks, such as the network alignment problem. In general, network alignment aims to discover a set of node pairs across two (or more) networks that we assume inherently have a correspondence between their nodes. The majority of existing network alignment algorithms assume additional constraints to guide the alignment process such as a one-to-one mapping between the two networks~\cite{zhang2015multiple}, some seed node-node correspondences (i.e., supervised)~\cite{mu2016slink}, sparsity in the possible alignments~\cite{bayati2013netalignbp}, and  the existence of side-information (e.g., node/edge attributes)~\cite{zhang2016final}. However, inherently these constraints limit the applications of these methods as in many cases these constraints are not available due to many reasons such as data privacy. Thus, this leaves the desire for an advanced algorithm that is both unsupervised and assuming no side-information, which brings in tremendous challenges. 

Without additional constraints, one key challenge to build network alignment algorithms is the vast number of possible permutations of the node orderings to align nodes from one network to another~\cite{heimann2018regal}. Previous works have focused on utilizing the adjacency matrix, or more recently, also leveraging spectral graph theory and the Laplacian matrix representations~\cite{nassar18lowrank,hayhoe2018spectral}. In these formulations,  the main idea is to discover the optimal permutation to map one network's matrix representation to that of the other with minimal variation between them.  Various metrics have been defined to measure the similarity between these matrices during the optimization process~\cite{guzzi2017survey,aflalo2015convex}. Inherently the use of the adjacency matrix is not scalable. Recently though, the field of network embedding, which in general seeks to discover a low dimensional representation of the nodes in a network, has seen amazingly fast development with advanced methods providing huge improvement over purely spectral based methods for single network tasks~\cite{grover2016node2vec}. This is primarily due to the condensed, space efficient, and even richer low dimensional representations for the nodes of a network. However, these network embedding methods are optimized separately for different graphs. In other words, embeddings of nodes from two networks are not aligned. Thus, directly applying network embedding to advance network alignment still is immensely challenging. 

Meanwhile, there have been adversarial based methods~\cite{goodfellow2014gan,isola2017visiongan,yu2017nlpgan,wang2017visiongan} that harness the power of deep learning for solving a variety of unsupervised problems by using a minimax game between a generator and a discriminator. In these adversarial based methods, the generator is trained to attempt at ``fooling'' the discriminator that it is generating ``real'' (and not ``generated'') examples while the discriminator is also trained to get better at differentiating between the ``real'' and ``generated'' examples. This process allows for an unsupervised way of learning a generator that can generate examples that seemingly come from the same distribution of the real data. These adversarial techniques have shown to be useful in a plethora of domains including computer vision~\cite{isola2017visiongan}, natural language processing~\cite{yu2017nlpgan}, and recommendation~\cite{wang2017visiongan}. 

On the one hand, network embedding algorithms have been proven to be effective in learning representations for nodes, but embeddings for two networks are learned separately, which are not aligned. On the other hand, adversarial techniques are powerful in learning real data distributions. Thus, in this work, we propose to harness the power of network embedding and adversarial techniques to tackle the challenging network alignment problem without additional constraints or knowledge outside of the network structure. The rationale is that we can align the node representations of two networks by taking advantage of adversarial techniques. More specifically, the proposed novel Deep Adversarial Network Alignment (DANA) framework is composed of two stages -- one graph distribution alignment stage and one node alignment stage. In the graph distribution alignment stage,  we utilize deep neural networks in an adversarial framework that is able to learn a highly complex mapping from one network's embedding space to that of the other such that the mapped embedding approximates the data distribution of the other network's original embedding. In the node alignment stage, we align individual nodes from two networks by using the mapping functions learned from the graph distribution alignment stage. Our main contributions are as follows:
\begin{itemize} 
\item We propose a novel unsupervised Deep Adversarial Network Alignment (DANA) framework that utilizes the power of both network embedding and adversarial training techniques to align the embedding distributions and then perform an efficient node alignment thereafter;
\item We provide an unsupervised heuristic to perform model selection for DANA, which also simultaneously shows the effectiveness of aligning the embedding distributions; and 
\item Experimental results on various datasets show the superiority of DANA against numerous advanced baselines.
\end{itemize}
\section{Problem Definition} \label{sec:problem}

In this section, we introduce the basic notations and problem definition. 
First, we let $N^1=(\mathcal{V}^1,\mathcal{E}^1)$ and $N^2=(\mathcal{V}^2,\mathcal{E}^2)$ be two undirected networks with $\mathcal{V}^1 = \{v^1_1, v^1_2, \dots v^1_{n_1} \}$ and $\mathcal{V}^2 = \{v^2_1, v^2_2, \dots v^2_{n_2} \}$ being sets of $n_1$ and $n_2$ vertices, and edge sets $\mathcal{E}^1$ and $\mathcal{E}^2$ for networks $N^1$ and $N^2$, respectively. 

Now, with the aforementioned notations, we formally define the network alignment problem we want to study in this work as follows:

\textit{Given two networks $N^1$ and $N^2$, and under the assumption that there is an underlying correspondence between the vertices $\mathcal{V}^1$ and $\mathcal{V}^2$, we seek to discover a set $\mathcal{A}$ of vertex alignment pairs defined as:}
\begin{align}\small 
\mathcal{A} = \{(v^1,v^2) ~|~  v^2 \in \mathcal{V}^2 , \forall v^1 \in \mathcal{V}^1\} 
\end{align}
\textit{where for each vertex $v^1$ in $N^1$ we predict a single corresponding vertex $v^2$ in $N^2$, such that together these pairwise node alignments follows a global network alignment.} 

Actually we will solve this problem bidirectionally to align the nodes of $N^1$ to $N^2$ and vice versa. Furthermore, we stress that in our unsupervised setting, we do not have any known node-node labeled correspondences nor any side-information (such as node/edge attributes). Instead, our proposed framework only requires the network structures, but could be extended to embrace such additional information (later discussed as future work).    
\section{Deep Adversarial Network Alignment Framework} \label{sec:framework}

In this section we introduce our proposed framework, Deep Adversarial Network Alignment (DANA), for the network alignment problem discussed in Section~\ref{sec:problem}. First, we will provide an overview of how the framework is utilized to solve the network alignment problem by first aligning the embedding distributions and then aligning the nodes. Next, we will discuss in detail both of these key stages of our proposed framework. Thereafter we summarize with an algorithmic overview of our framework and also provide an analysis on the complexity of DANA.

As previously mentioned, network embedding algorithms have been proven to effectively learn node representations, but embeddings for separate networks are not aligned. Thus, we first obtain node embeddings and then use an adversarial based method to correctly learn a complex (and even non-linear) mapping to simultaneously align the two networks embedding distributions. In this way the mapped embedding from the first network approximately follows the distribution of the other. Then, once the distributions have been aligned the second stage  uses an efficient nearest neighbor search to match/align the individual nodes by using the mapping functions obtained through the adversarial learning.

\begin{figure}[t]
\begin{center}
\includegraphics[width=0.95\columnwidth]{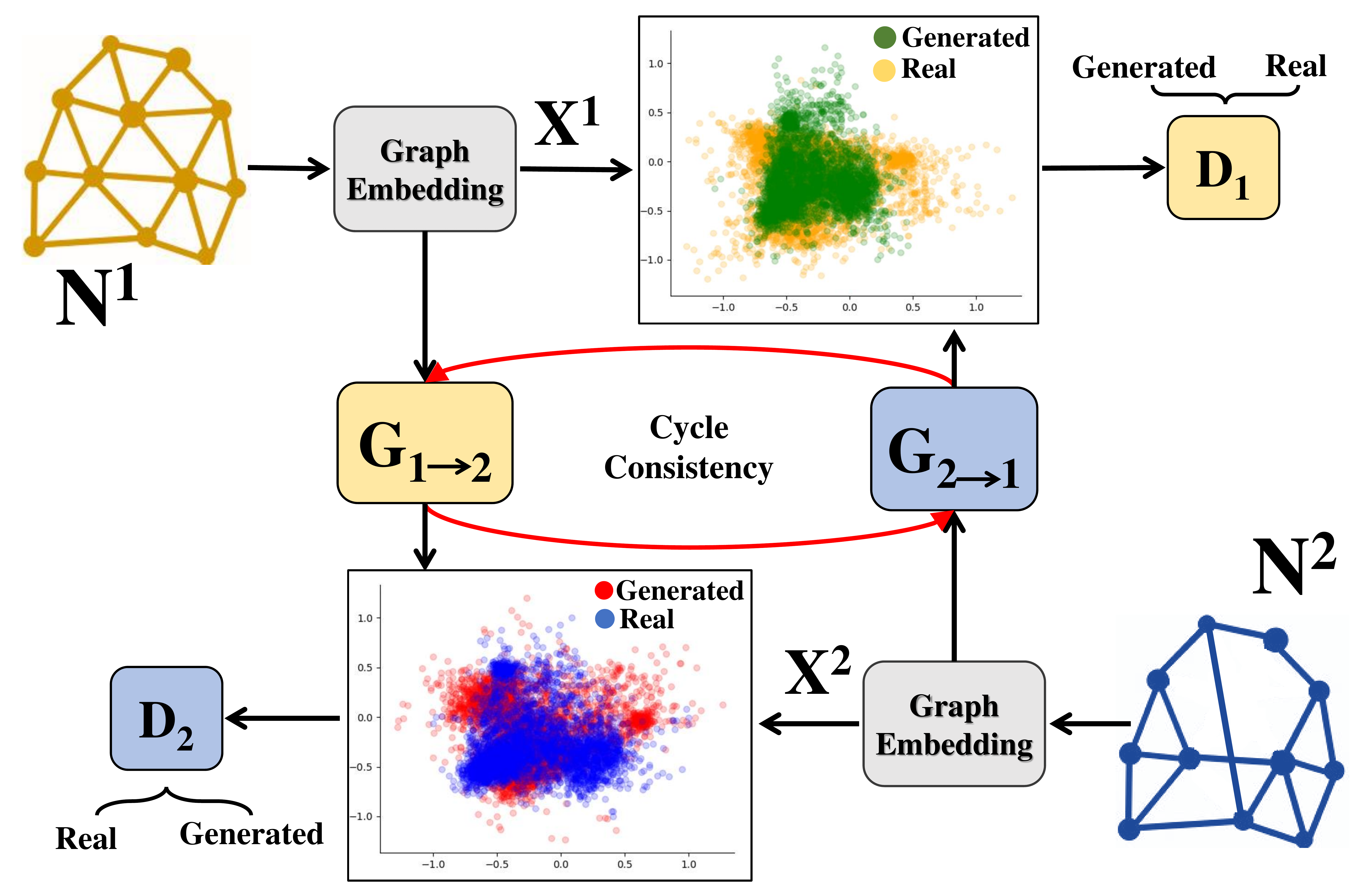}
\end{center}
\caption{An illustration of our graph distribution alignment.}
\label{fig:model}
\vskip -1ex
\end{figure}

\subsection{Adversarial Graph Distribution Alignment}
\label{sec:gan}
In this section, we introduce the first stage of DANA, namely the distribution alignment whose model is illustrated in Figure~\ref{fig:model}. In a nutshell, this model aligns two graphs $N^1$ and $N^2$ bidirectionally using two connected adversarial networks. The reason for connecting them is to ensure no ``collapse''~\cite{zhu2017cyclegan} and utilize transitivity~\cite{zhou2016cycleconsistency} to regularize and prevent random alignments of the distributions which could be possible due to using complex (and even non-linear) mappings between them. 

In Figure~\ref{fig:model}, we can observe that when given two networks $N^1$ and $N^2$ the first step is to obtain graph embeddings for each of these networks. This can be obtained using one of the plethora of available methodologies, such as node2vec~\cite{grover2016node2vec}. Then, our goal is to find a mapping between node embeddings $\mathbf{X}^{1}=\{x^1_i\}_{i=1}^{|\mathcal{V}^1|}$ for $N^1$ and  $\mathbf{X}^{2}=\{x^2_j\}_{j=1}^{|\mathcal{V}^2|}$ for $N^2$ whose distributions are denoted as $x^1 \sim p_{emb}(x^1) $ and $x^2 \sim p_{emb}(x^2) $, respectively (since we are ultimately looking to align the two graph distributions). As demonstrated in Figure~\ref{fig:model}, the model contains two generators $G_{1 \rightarrow 2}$ mapping $\mathbf{X}^1 \rightarrow \mathbf{X}^2 $ and $G_{2 \rightarrow 1}$ mapping $\mathbf{X}^2 \rightarrow \mathbf{X}^1$. Moreover, the discriminator $D_{1}$ distinguishes between real embeddings of graph $N^1$ and those generated from the real embedding of $N^2$ through $G_{2 \rightarrow 1}$ (i.e., $\{x^1\}$ and $\{ G_{2 \rightarrow 1}(x^2) \}$, respectively). Likewise, the discriminator $D_{2}$ distinguishes between real embeddings of $N^1$ and those generated through $G_{1 \rightarrow 2}$ from the real embedding of $N^2$. 

Both bidirectional mappings are optimized via applying adversarial losses~\cite{goodfellow2014gan}. More precisely, the loss function associated with aligning graph $N^1$ to $N^2$ (i.e., the mapping $G_{1 \rightarrow 2}: \mathbf{X}^1 \rightarrow \mathbf{X}^2$) is as follows:
\begin{align}\label{eq:gan1}
  \min_{G_{1 \rightarrow 2}} \max_{D_2}~ &\mathcal{L}^{1 \rightarrow 2}_{adv}(G_{1 \rightarrow 2},D_{2},\mathbf{X}^1,\mathbf{X}^2) \\  
    &= \mathbb{E}_{ x^2 \sim p_{emb}(x^2)} [log~ D_{2}(x^2)] \nonumber \\
    &+ \mathbb{E}_{ x^1 \sim p_{emb}(x^1)} [log\; \big( 1-D_{2}(G_{1 \rightarrow 2}( x^1)) \big) ] \nonumber  
\end{align}

\noindent where the generator $G_{1 \rightarrow 2}$ is trying to mimic the embedding distribution of $N^2$ by generating embeddings $G_{1 \rightarrow 2}(x^1)$ (by minimizing Eq.~(\ref{eq:gan1})) while simultaneously the discriminator $D_2$ is attempting to differentiate between $x^2$ and $G_{1 \rightarrow 2}(x^1)$ (by maximizing Eq.~(\ref{eq:gan1})). 
Similarly, the loss function of aligning graph $N^2$ to $N^1$  (i.e., $G_{2 \rightarrow 1}: \mathbf{X}^2 \rightarrow \mathbf{X}^1$) is as follows: 
\begin{align}\label{eq:gan2}
  \min_{G_{2 \rightarrow 1}} \max_{D_1}~ &\mathcal{L}^{2\rightarrow1}_{adv}(G_{2 \rightarrow 1},D_{1},\mathbf{X}^1,\mathbf{X}^2) \\  
    &= \mathbb{E}_{ x^1 \sim p_{emb}(x^1)} [log\; D_{1}(x^1)] \nonumber \\
    &+ \mathbb{E}_{ x^2 \sim p_{emb}(x^2)} [log\; \big( 1-D_{1}(G_{2 \rightarrow 1}( x^2)) \big)] \nonumber  
\end{align}
\noindent where in this situation the minimax game is instead between $G_{2 \rightarrow 1}$ and $D_1$. 

By separately optimizing the loss functions in Eq.~(\ref{eq:gan1}) and Eq.~(\ref{eq:gan2}) (i.e., of learning the mappings $G_{1 \rightarrow 2}$ and $G_{2 \rightarrow 1}$), we might expect to learn an alignment between the embeddings of $N^1$ to $N^2$, and vice versa. However, in practice, during the training each of the separate models can map one real embedding distribution to some random embeddings in the target domain (or even collapse). More specifically, especially when non-linearity is used in the generators, the mapping $G_{1 \rightarrow 2}$ can project embeddings of graph $N^1$ to some random points in the embedding space of $N^2$, that although might have a similar distribution, might also have completely distorted the proximity information between neighboring nodes that was originally preserved in ${\bf X}^1$. In other domains, such as word-to-word translation, adversarial techniques resorted to using only a single directional linear mapping~\cite{lample2018wordgan} that could avoid these problems, but limited the complexity and power of non-linearity in their translation/alignment. 
 
To prevent these problems and still potentially use the power of a non-linear generator mapping function for network alignment, we introduce the cycle consistency loss similar to~\cite{zhu2017cyclegan}, which had been used for image-to-image translation.
More specifically, for a node embedding $x^1$ of node $v^1 \in \mathcal{V}^1$, the learned generators $G_{1 \rightarrow 2}$ and $G_{2 \rightarrow 1}$ should be able to recover and bring $x^1$ back to the embedding space of $N^1$ as follows:
\begin{align}
x^1 \rightarrow G_{1 \rightarrow 2}(x^1) \rightarrow  G_{2 \rightarrow 1}(G_{1 \rightarrow 2}(x^1)) \approx x^1
\end{align}
and similarly for being able to recover the embeddings of $N^2$. Intuitively, if forcing these cyclic mappings, then this would help to prevent both the ``collapse'' and random alignment problem previously mentioned. Hence, we incorporate the following cycle-reconstruction loss into our objective:
\begin{align}    \label{eq:lossconsitency}
        &\min_{G_{2 \rightarrow 1},G_{1 \rightarrow 2}} \mathcal{L}_{cyc}(G_{1 \rightarrow 2}, G_{2 \rightarrow 1},\mathbf{X}^1,\mathbf{X}^2) \\
        & = \mathbb{E}_{x^1 \in p_{emb}(x^1)}\Big[||G_{2 \rightarrow 1} \big (G_{1 \rightarrow 2}(x^1) \big ) -x^1 ||_{1}\Big] \nonumber \\ 
        & + \mathbb{E}_{x^2 \in p_{emb}(x^2)}\Big[||G_{1 \rightarrow 2} \big (G_{2 \rightarrow 1}(x^2) \big ) -x^2 ||_{1}\Big] \nonumber 
\end{align}

This leads to the graph level embedding distribution alignment to optimize the following overall loss function:
\begin{align}    \label{eq:totalloss}
        &\min_{G_{1 \rightarrow 2},G_{2 \rightarrow 1}} \max_{D_1,D_2} \mathcal{L}(G_{1 \rightarrow 2},G_{2 \rightarrow 1},D_1,D_2,{\bf X}^1,{\bf X}^2) \\
        &=  \mathcal{L}^{1 \rightarrow 2}_{adv}(G_{1 \rightarrow 2},D_{2} , \mathbf{X}^1 , \mathbf{X}^2) + \mathcal{L}^{2 \rightarrow 1}_{adv}(G_{2 \rightarrow 1},D_{1},\mathbf{X}^1,\mathbf{X}^2) \nonumber \\ 
        & + \lambda  \mathcal{L}_{cyc}(G_{1 \rightarrow 2}, G_{2 \rightarrow 1},{\bf X}^1,{\bf X}^2) \nonumber 
\end{align}

\noindent where $\lambda$ is a hyper-parameter controlling the balance between ensuring a close aligning of the graph level embedding distributions and the cycle consistency loss. 

\begin{figure}
    \centering
    \includegraphics[scale=0.2]{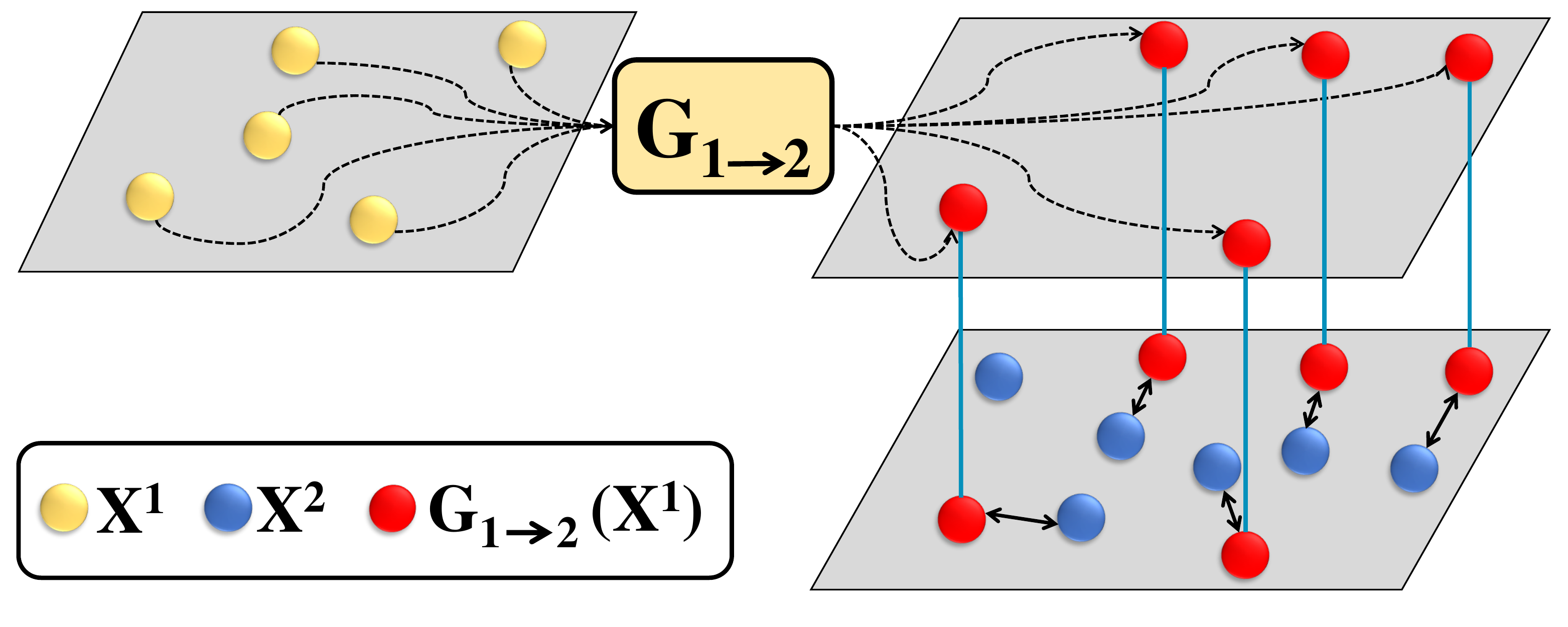}
    \caption{An illustration of our node alignment from $N^1$ to $N^2$.}
    \label{fig:3DIllustration}
    \vskip -1ex
\end{figure}

\subsection{Nearest Neighbor Node Alignment}
The second stage of DANA is to efficiently discover the node alignments, which are based on using the discovered complex mapping functions (i.e., the generators) from the first stage of DANA. In this subsection, we will discuss the efficient nearest neighbor greedy node alignment method from $N^1$ to $N^2$, where we can then similarly perform $N^2$ to $N^1$.

As seen in Figure~\ref{fig:3DIllustration}, the first step is to take the node embeddings ${\bf X}^1$ and map them to the embedding space of $N^2$ through the use of the trained generator $G_{1 \rightarrow 2}$. Next these projected node embeddings of $N^1$ are paired with their nearest neighbor from $N^2$ based on Euclidean distance. To perform the nearest neighbor search, we utilize a k-d tree, which is a data structure used for performing a fast and efficient search.~\cite{abbasifard2014search}.

\subsection{\hspace{-1ex}Algorithmic Overview and Complexity Analysis}\label{sec:overview}
Here we discuss an algorithmic overview of DANA along with the computational complexity. Algorithm~\ref{alg:DANA} summarizes the entire framework including the major steps-- namely obtaining network embeddings (line 2), training the unsupervised adversarial based graph distribution alignment (lines 4-11), and performing nearest neighbor node alignment (lines 13-15). Next we discuss some details and the complexity of DANA. Note that we denote $n_{max} = max(n_1,n_2)$, and similarly define $n_{min}$, where $n_1 = |\mathcal{V}^1|$ and $n_2 = |\mathcal{V}^2|$.

First, we obtain the embeddings for networks $N^1$ and $N^2$. In this work, we utilize a network embedding method (more specifically node2vec~\cite{grover2016node2vec}) whose complexity is $O(n_1)$ and $O(n_2)$ for networks $N^1$ and $N^2$, respectively, resulting in overall $O(n_{max})$. Note that DANA could use attributed embedding methods to incorporate side-information, but we leave this as future work.

Next, we train the adversarial based graph distribution alignment.
Suppose that the algorithm is run for some constant number of epochs, $K$, where each epoch iterates through all the nodes (for both graphs) by randomly creating mini-batches that perform the forward step, backpropagating error and also updating the parameters using stochastic gradient descent (SGD). Note that this depends on the architectures for the generators and discriminators. However, as later discussed in Section~\ref{sec:experiments}, if we use small reasonable constant size hidden layers, then the computation is also constant for each mini-batch. Thus, due to the fact we run $K$ epochs, the complexity of the graph distribution alignment is $O(n_{max})$.

As seen in Algorithm~\ref{alg:DANA} (lines 13-15), for the final step we actually perform the node alignment bidirectionally and select the one that has the lower average nearest neighbor distance. First, we build the k-d tree based on the embeddings $\mathbf{X}^2$, which takes $O(n_2 log(n_2))$ and then we need to search for the nearest neighbor for all $v^1 \in \mathcal{V}^1$ using their mapped representation $G_{1 \rightarrow 2}(x^1)$. The search in the worse case for each $x^1$ is $O(n_2)$, but $O(log(n_2))$ in the average case. Thus, since we perform this search for all nodes in $\mathcal{V}^1$, the total expected time is $O(n_2 log(n_2) + n_1 log(n_2))$ for the alignment of $N^1$ onto $N^2$. Then we similarly do the alignment of $N^2$ onto $N^1$, resulting in the total expected time $O(n_{max} log(n_{max}))$. This leads to the sub-quadratic total time complexity of DANA to be $O(n_{max} log(n_{max}))$ when ignoring the linear/constant terms. 

\begin{algorithm}[t]
\footnotesize
 \DontPrintSemicolon
 \KwIn{$\mathcal{N}^1=(\mathcal{V}^1,\mathcal{E}^1), \mathcal{N}^2=(\mathcal{V}^2,\mathcal{E}^2)$}
 \KwOut{$\mathcal{A}$}
 \textbf{\# Utilize a Graph Embedding Method:}\;
  Obtain $\mathbf{X}^1$ and  $\mathbf{X}^2$ from $N^1$ and $N^2$, respectively\;   
 
 \textbf{\# Perform Graph Distribution Alignment}\;
 Randomly initialize the neural network parameters ($\theta$)\\
 \While{Not max iterations}
 {
 
    Select a random mini-batch $\mathcal{B} = \{(x^{1},x^{2}) | x^{1} \in \mathbf{X}^1,  x^{2} \in \mathbf{X}^2 \}$  
    
    \ForEach{ $(x^{1},x^2) \in \mathcal{B}$}
    {
        Feed forward to obtain $G_{1 \rightarrow 2}(x^1)$, $G_{1 \rightarrow 2}(x^2)$, $D_2(G_{1 \rightarrow 2}(x^1))$, $D_1(G_{2 \rightarrow 1}(x^2))$, $D_1(x_1)$, $D_2(x^2)$, $G_{2 \rightarrow 1}(G_{1 \rightarrow 2}(x^1))$, and $G_{1 \rightarrow 2}(G_{2 \rightarrow 1}(x^2))$
    }
  Update parameters $\theta$ in Eq.~\ref{eq:totalloss} using SGD\\
 }
\textbf{\# Perform Node Alignment}\;
   Construct $\mathcal{A}^1$: closest $x^2 \in$ k-d\_tree($\mathbf{X^2})$ $\forall x^1 \in G_{1 \rightarrow 2}(\mathbf{X}^1)$
   
   Construct $\mathcal{A}^2$: closest $x^1 \in$ k-d\_tree($\mathbf{X^1})$ $\forall x^2 \in  G_{2 \rightarrow 1}(\mathbf{X}^2)$
    
   Set $\mathcal{A}$ to $\mathcal{A}^1$ or $\mathcal{A}^2$ based on mean nearest neighbor distance
    
\caption{Deep Adversarial Network Alignment.}
\label{alg:DANA}
\end{algorithm} 
\section{Experiments}\label{sec:experiments}
To evaluate the effectiveness of the proposed Deep Adversarial Network Alignment framework, we conduct a set of experiments for aligning real world networks. Through the conducted experiments, we seek to answer the following two questions: (1) Can DANA align network embeddings? and (2) How effective is DANA at accurately discovering the true underlying node alignment? 

\subsection{Experimental Setup}
Here we will discuss the datasets and how we utilize them for our experiments, the architecture details for DANA, our proposed unsupervised heuristic for model selection, and baseline methods.

\subsubsection{Datasets with Ground Truth Correspondence}
The first two datasets we collected are Bitcoin-Alpha\footnote{http://www.btcalpha.com} and Bitcoin-OTC\footnote{http://www.bitcoin-otc.com}. These networks are online marketplaces that allow users to buy and sell things using Bitcoins. Users create positive (or negative) links to those they trust (or distrust). Furthermore, most users have provided their unique Bitcoin fingerprints, thus allowing us to determine a ground-truth mapping of users across networks, which we use to evaluate the alignments. Note that we construct two undirected dataset variants, the first being networks that only include positive links (i.e., BitcoinA and BitcoinO), and the second that also include the negative links (i.e., BitcoinAn and BitcoinOn). Some basic statistics can be found in Table~\ref{tab:datastatistics}. 

\subsubsection{Datasets with Pseudo-Ground Truth Correspondence}
Here we collected real world datasets, namely CollegeMsg\footnote{http://snap.stanford.edu/data/}, Hamsterster\footref{konect}, and Blogs\footnote{\label{konect}http://konect.uni-koblenz.de/}. We present the basic undirected network statistics in Table~\ref{tab:datastatistics}. Note that we have chosen these datasets to be used in a synthetic network alignment setting (although themselves are real world networks), where we will let $N_2$ to be the original dataset, while constructing a permuted version as $N_1$ and simultaneously adding some noise in the set \{5,10,20\}\% to $N_1$ at random to evaluate the performance and robustness of DANA. 

\begin{table}[]
    \centering
    \footnotesize
    \setlength\tabcolsep{2.5pt}
    \begin{tabular}{|c|c|c|c|c|c|}
    \hline
        &\makecell{ BitcoinA \\ BitcoinO \\ (pos)} & \makecell{ BitcoinAn \\  BitcoinOn \\ (pos\&neg)} & CollegeMsg &  Hamsterster & Blogs \\ \hline 
        
        $|\mathcal{V}^1|$ & 3682 & 3783 & 1899 & 2426 & 1224   \\
        
        $|\mathcal{V}^2|$ & 3819 & 3914 & - & - & -\\
        
        $|\mathcal{V}^1 \cap \mathcal{V}^2|$ & 3591 & 3682 & - & - & -\\ \hline 
        
        $|\mathcal{E}^1|$ & 25952 &  28288 & 13754 & 16613 & 16718 \\
        
        $|\mathcal{E}^2|$ & 28321 & 30691 & - & - & -\\
        
        $|\mathcal{E}^1 \cap \mathcal{E}^2|$ & 24066 & 26004  & - & - & -\\ \hline 
    \end{tabular}
    \caption{Dataset Statistics.}
    \label{tab:datastatistics}
    \vskip -3ex
\end{table}

\subsubsection{DANA Architecture} 
First, for the graph embeddings, we utilize node2vec~\cite{grover2016node2vec} to obtain node embeddings of size 64. We note that both the generator and discriminator for DANA can be constructed in various ways. For the discriminator, based on knowledge from other domains when using adversarial based frameworks, we use a two layer fully connected network with 512 hidden units and Leaky ReLU (Rectified Linear Unit)~\cite{maas2013rectifier}. For the generators, we attempted using both a linear and non-linear single layer mapping (i.e., two possible variants). For the adversarial learning, we varied $\lambda$ in \{1, 10, 100\}, since $\lambda = 10$ was recommended in~\cite{zhu2017cyclegan}. We let $\eta \in \{1,5,25\}$ denote the number of times we update the generators before updating the discriminators during the alternative updating and use the ADAM optimizer~\cite{kingma2014adam}.

In the above we mentioned three hyperparameters that we would need to choose between, thus, we propose to use the unsupervised heuristic of how well DANA aligns the distributions as a metric to select the best parameters. We assume the better the embedding distributions match, the better the performance in aligning the individual nodes. \textit{Note that this does not utilize the ground-truth alignments, but rather simply measures the average distance each mapped node is to the nearest neighbor in the other embedded space}. 

\subsubsection{Baselines} 
Here we introduce the set of baselines we will compare against our proposed Deep Adversarial Network Alignment (DANA) method. Isorank~\cite{singh2008isorank}/FINAL~\cite{zhang2016final} is a network alignment algorithm that was designed specifically for protein-protein interaction network alignments and we note that FINAL is an attributed network alignment methods that is equivalent to IsoRank when having no edge/node attributes~\cite{zhang2016final}. EigenAlignLR~\cite{nassar18lowrank} is a low rank extension of the EigenAlign~\cite{feizi2016spectral} spectral based network alignment method. REGAL~\cite{heimann2018regal} constructs their own network embedding (while REGAL-s2v~\cite{heimann2018regal} instead uses struc2vec~\cite{ribeiro2017struc2vec}) and then uses a nearest neighbor search for node alignments. We also include two sparse network alignment methods, namely SparseIsoRank
~\cite{bayati2013netalignbp} (a sparse network alignment variation of IsoRank~\cite{singh2008isorank}) and NetAlignBP~\cite{bayati2013netalignbp} (which uses belief propagation to construct the alignment), where both use information for limiting the scope of possible alignments.

We note that SparseIsoRank and NetAlignBP both assume additional information to suggest to their methods which node alignments are possible, but our problem setting does not have such additional information. Therefore we heuristically provide them information from the network structure. More specifically, node degree similarity from $N^1$ and $N^2$ is used to provide each node in the two networks a subset of possible nodes to pair with (as also done in~\cite{heimann2018regal}). Here we try using $log_2(|\mathcal{V}|)$ and $0.01 \times |\mathcal{V}|$ for the the set size containing the most similar nodes in terms of absolute difference in degree. We note that for all methods we use the default settings provided by the authors, but REGAL was unable to run on networks of different sizes, thus we only report their performance for the CollegeMsg, Blogs, and Hamsterster datasets. 

\begin{figure}[]
\begin{center}
\includegraphics[scale=0.28]{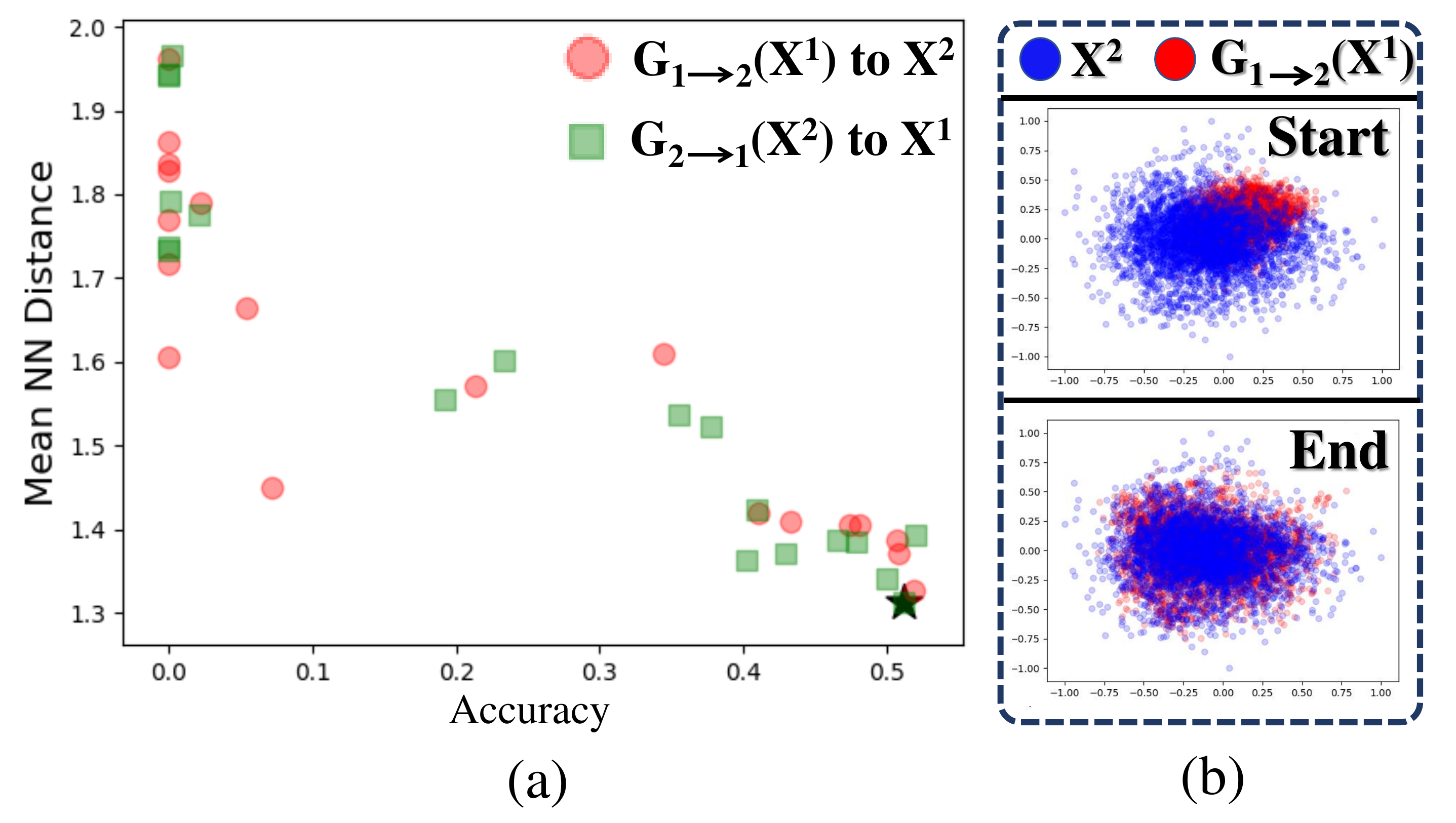}
\end{center}
\vspace{-2ex}
\caption{Answering question (1) where (a) shows the average nearest neighbor distance heuristic with the accuracy for BitcoinAn and BitcoinOn, and (b) visualizes the graph distribution  alignment.} 
\label{fig:correlation}
\vskip -1ex
\end{figure}

\subsection{Experimental Results}
Here we present the results of our experiments in network alignment with both our known ground truth and pseudo-ground truth datasets. Although there are multiple ways of evaluating the performance of network alignment methods~\cite{douglas18metrics}, we report the accuracy, which is by far the most commonly used. 

In Figure~\ref{fig:correlation}(a), we have plotted the mean nearest neighbor distances against the accuracy while taking snapshots during DANA's optimization of the adversarial graph distribution alignment of the BitcoinAn and BitcoinOn datasets. We can observe that as DANA better aligns the two distributions (i.e., lower mean nearest neighbor distance) the accuracy is also improving. In Figure~\ref{fig:correlation}(a), the star represents the least mean nearest neighbor distance and we can observe it nearly has the best accuracy. We observe the same trend across all hyper-parameter settings and thus our unsupervised heuristic of using the lowest mean nearest neighbor distance for model selection works quite well in practice. 
To further show the effectiveness of DANA in aligning the graph embedding distributions, we show a visualization using principle component analysis (PCA)~\cite{jolliffe2011principal} in Figure~\ref{fig:correlation}(b) where ``Start'' refers to the initial random alignment and ``End'' shows the final graph distribution alignment that DANA adversarially learns. Therefore, based on Figure~\ref{fig:correlation}, we can conclude an answer for our first question, that DANA can indeed learn to effectively align the embeddings of two networks.

In Table~\ref{tab:bitcoints}, we ran DANA and the baselines (except for REGAL, since their implementation could not handle networks having different sizes) on the Bitcoin datasets. The first observation is that the EigenAlignLR, SparseIsoRank, and Isorank/FIANL all have significantly less performance than NetAlignBP, REGAL-s2v, and DANA. It would seem that NetAlignBP is able to effectively use the pseudo side-information (based on node degree similarity) we provided. Also, REGAL-s2v (which uses struc2vec embeddings) is able to out perform EigenAlignLR (which is spectral based). However, we also note the comparison of REGAL-s2v, which directly performs a node alignment on embeddings from struc2vec, against DANA, which uses adversarial learning to correctly align network embeddings before performing the node alignment. We can clearly see that DANA significantly out performs REGAL-s2v and all other baseline methods. 

\begin{table}\small 
    \centering
    \footnotesize 
    \begin{tabular}{|c|c|c|}
    \hline
        Methods  & \makecell{ $N^1$: BitcoinA \\ $N^2$: BitcoinO} & \makecell{ $N^1$: BitcoinAn \\ $N^2$: BitcoinOn}  \\ \hline\hline 
        SparseIsoRank & 0.046 & 0.047 \\ \hline
        NetAlignBP & 0.157 & 0.141 \\ \hline
        IsoRank/FINAL &0.041 &0.040 \\ \hline
        EigenAlignLR & 0.015 &0.016 \\ \hline
        REGAL-s2v & 0.124 & 0.089 \\ \hline
       DANA &\textbf{0.542} & \textbf{0.511}\\ \hline
    \end{tabular}
    \vspace{-1ex}
    \caption{Performance comparison with accuracy for aligning the two variations of the Bitcoin datasets.}
    \label{tab:bitcoints}
    \vspace{-2ex}
\end{table}

\begin{figure}[ht]
\hspace*{-5mm}    
\includegraphics[scale=0.26]{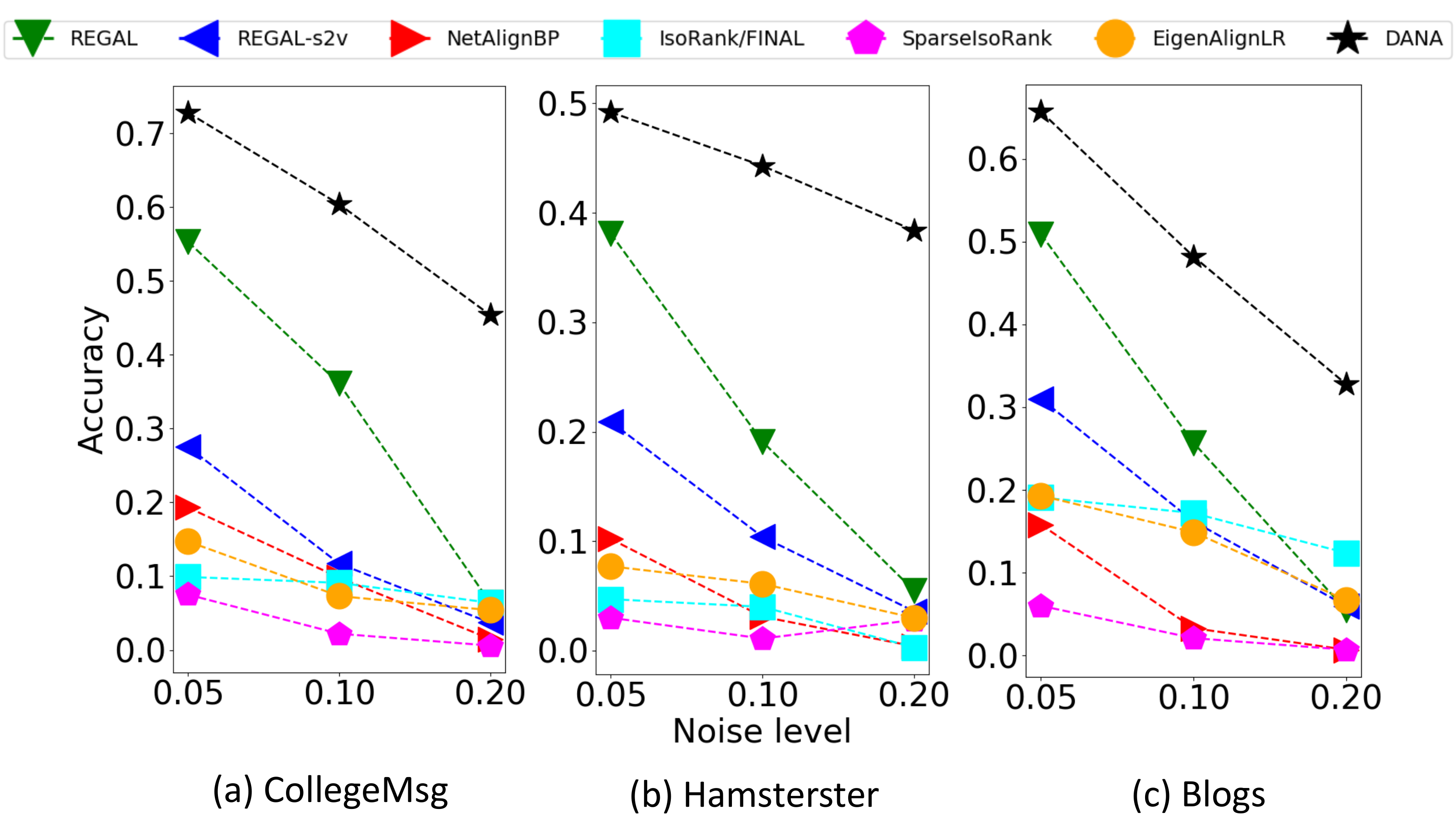}
    \caption{Results on the three datasets with pseudo-ground truth when varying the level of noise (i.e., percent of edges removed).}
    \label{fig:pseudodatasets}
    \vskip -1ex
\end{figure}

Next, in Figure~\ref{fig:pseudodatasets}, we present the results for the pseudo-ground truth datasets where we have performed the network alignment experiments for the CollegeMsg, Hamsterster, and Blogs datasets. In these experiments we first permuted the nodes of the original network and then removed a portion of the edges (i.e., level of noise) and attempted to align back to the original network. We can observe that similar to the two Bitcoin experiments, DANA is able to outperform all existing baseline methods for all three datasets across all levels of noise. We also observe that as more edges are removed, it becomes harder to align back to the original network where all the baselines almost completely fail to align at 20\%, but yet DANA is able to still maintain a reasonable alignment.
It can also be seen that DANA and REGAL (also REGAL-s2v in many cases) outperform the other methods, which suggests again that embedding based approaches are superior. However, as previously mentioned, REGAL-s2v does not perform any alignment of the embeddings before performing the node alignment, and REGAL performs a similarity-based embedding alignment through a shallow matrix factorization method, but neither harness deep learning or adversarial training. Furthermore, while EigenAlignLR uses spectral based embeddings, DANA is able to harness more advanced network embedding methods, while leading to better performance across all levels of noise. This seems natural due to the fact that current network embedding methods have shown superiority over the classical spectral embeddings for a variety of network analysis tasks~\cite{grover2016node2vec}.  Thus, based on these experiments we have answered the second question, DANA is indeed effective at aligning the corresponding nodes across networks, which is due to the fact it harnesses the power of deep adversarial learning to correctly align the embeddings of two networks. 

\section{Related Work} 
Network alignment is a fundamental network analysis task having many real world applications in user-identity linkage~\cite{liu2013ulinkage}, computer vision~\cite{conte2004survey}, and bioinformatics~\cite{singh2008isorank}.  Classical network alignment methodologies typically were based around optimizing a permutation matrix to align the matrix representations. However, some methods have introduced relaxations such as convex or finding a doubly stochastic matrix instead of finding a permutation matrix~\cite{aflalo2015convex}.


Another set of related network alignment problems are those that are supervised. Some representative examples also learn network embeddings, but use known node-node pairs to align in a shared embedding space~\cite{tan2014elinkage}. 
Also, there is the sparse network alignment problem~\cite{bayati2013netalignbp} where the general alignment problem is simplified to restrict the possible alignments between nodes. In other words, a bipartite graph is constructed from the two networks being aligned, but rather than having $|\mathcal{V}^1| \times |\mathcal{V}^2|$ number of possible matching pairs, they instead have a limited set to prevent certain pairs, which could be from domain specific knowledge or network structure. Some other specialized formulations can be found for 
heterogeneous networks~\cite{kong2013heterogeneous} and attributed graphs~\cite{zhang2016final}.


\section{Conclusion}

In this work, we proposed our Deep Adversarial Network Alignment (DANA) framework to solve the general network alignment problem when only provided the network structure and assuming no additional constraints. More specifically, DANA harnesses the power of adversarial learning to align the graph embedding distributions and then thereafter performs an efficient nearest neighbor node alignment. Furthermore, we present an unsupervised heuristic to perform model selection for DANA. Finally extensive experiments were performed to show the effectiveness of both main stages of DANA, while also proving DANA to be superior in performance against existing network alignment methods. 
Our future work consists of first extending DANA to embrace additional constraints to aid in performing alignments, such as node/edge attributes or assuming a seed set of known node-node aligned pairs. 

 
\clearpage
\newpage

\bibliographystyle{named}
\bibliography{references/network-alignment}

\begin{thebibliography}{}

\bibitem[\protect\citeauthoryear{Abbasifard \bgroup \em et al.\egroup
  }{2014}]{abbasifard2014search}
Mohammad~Reza Abbasifard, Bijan Ghahremani, and Hassan Naderi.
\newblock Article: A survey on nearest neighbor search methods.
\newblock {\em IJCA}, 2014.

\bibitem[\protect\citeauthoryear{Aflalo \bgroup \em et al.\egroup
  }{2015}]{aflalo2015convex}
Yonathan Aflalo, Alexander Bronstein, and Ron Kimmel.
\newblock On convex relaxation of graph isomorphism.
\newblock {\em PNAS}, 2015.

\bibitem[\protect\citeauthoryear{Bayati \bgroup \em et al.\egroup
  }{2013}]{bayati2013netalignbp}
Mohsen Bayati, David~F Gleich, Amin Saberi, and Ying Wang.
\newblock Message-passing algorithms for sparse network alignment.
\newblock {\em TKDD}, 2013.

\bibitem[\protect\citeauthoryear{Conte \bgroup \em et al.\egroup
  }{2004}]{conte2004survey}
Donatello Conte, Pasquale Foggia, Carlo Sansone, and Mario Vento.
\newblock Thirty years of graph matching in pattern recognition.
\newblock {\em IJPRAI}, 2004.

\bibitem[\protect\citeauthoryear{Douglas \bgroup \em et al.\egroup
  }{2018}]{douglas18metrics}
Joel Douglas, Ben Zimmerman, Alexei Kopylov, Jiejun Xu, Daniel Sussman, , and
  Vince Lyzinski.
\newblock Metrics for evaluating network alignment.
\newblock In {\em GTA3 at WSDM}, 2018.

\bibitem[\protect\citeauthoryear{Feizi \bgroup \em et al.\egroup
  }{2016}]{feizi2016spectral}
Soheil Feizi, Gerald Quon, Mariana Recamonde-Mendoza, Muriel M{\'e}dard,
  Manolis Kellis, and Ali Jadbabaie.
\newblock Spectral alignment of networks.
\newblock {\em arXiv preprint arXiv:1602.04181}, 2016.

\bibitem[\protect\citeauthoryear{Fortunato}{2010}]{fortunato2010commdetect}
Santo Fortunato.
\newblock Community detection in graphs.
\newblock {\em Physics Reports}, 2010.

\bibitem[\protect\citeauthoryear{Goodfellow \bgroup \em et al.\egroup
  }{2014}]{goodfellow2014gan}
Ian Goodfellow, Jean Pouget-Abadie, Mehdi Mirza, Bing Xu, David Warde-Farley,
  Sherjil Ozair, Aaron Courville, and Yoshua Bengio.
\newblock Generative adversarial nets.
\newblock In {\em NIPS}. 2014.

\bibitem[\protect\citeauthoryear{Grover and
  Leskovec}{2016}]{grover2016node2vec}
Aditya Grover and Jure Leskovec.
\newblock Node2vec: Scalable feature learning for networks.
\newblock SIGKDD, 2016.

\bibitem[\protect\citeauthoryear{Guzzi and
  Milenkovi{\'c}}{2017}]{guzzi2017survey}
Pietro~Hiram Guzzi and Tijana Milenkovi{\'c}.
\newblock Survey of local and global biological network alignment: the need to
  reconcile the two sides of the same coin.
\newblock {\em Briefings in bioinformatics}, 2017.

\bibitem[\protect\citeauthoryear{Hayhoe \bgroup \em et al.\egroup
  }{2018}]{hayhoe2018spectral}
Mikhail Hayhoe, Francisco Barreras, Hamed Hassani, and Victor~M Preciado.
\newblock Spectre: Seedless network alignment via spectral centralities.
\newblock {\em arXiv preprint arXiv:1811.01056}, 2018.

\bibitem[\protect\citeauthoryear{Heimann \bgroup \em et al.\egroup
  }{2018}]{heimann2018regal}
Mark Heimann, Haoming Shen, Tara Safavi, and Danai Koutra.
\newblock Regal: Representation learning-based graph alignment.
\newblock In {\em CIKM}, 2018.

\bibitem[\protect\citeauthoryear{Isola \bgroup \em et al.\egroup
  }{2017}]{isola2017visiongan}
P.~Isola, J.~Zhu, T.~Zhou, and A.~A. Efros.
\newblock Image-to-image translation with conditional adversarial networks.
\newblock In {\em CVPR}, 2017.

\bibitem[\protect\citeauthoryear{Jolliffe}{2011}]{jolliffe2011principal}
Ian Jolliffe.
\newblock Principal component analysis.
\newblock In {\em International encyclopedia of statistical science}. 2011.

\bibitem[\protect\citeauthoryear{Kingma and Ba}{2014}]{kingma2014adam}
Diederik~P Kingma and Jimmy Ba.
\newblock Adam: A method for stochastic optimization.
\newblock {\em arXiv preprint arXiv:1412.6980}, 2014.

\bibitem[\protect\citeauthoryear{Kong \bgroup \em et al.\egroup
  }{2013}]{kong2013heterogeneous}
Xiangnan Kong, Jiawei Zhang, and Philip~S. Yu.
\newblock Inferring anchor links across multiple heterogeneous social networks.
\newblock In {\em CIKM}, 2013.

\bibitem[\protect\citeauthoryear{Lample \bgroup \em et al.\egroup
  }{2018}]{lample2018wordgan}
Guillaume Lample, Alexis Conneau, Marc'Aurelio Ranzato, Ludovic Denoyer, and
  Hervé Jégou.
\newblock Word translation without parallel data.
\newblock In {\em ICLR}, 2018.

\bibitem[\protect\citeauthoryear{Liben-Nowell and
  Kleinberg}{2007}]{liben-nowell2007linkpred}
David Liben-Nowell and Jon Kleinberg.
\newblock The link-prediction problem for social networks.
\newblock {\em JAIST}, 2007.

\bibitem[\protect\citeauthoryear{Liu \bgroup \em et al.\egroup
  }{2013}]{liu2013ulinkage}
Jing Liu, Fan Zhang, Xinying Song, Young-In Song, Chin-Yew Lin, and Hsiao-Wuen
  Hon.
\newblock What's in a name?: An unsupervised approach to link users across
  communities.
\newblock WSDM, 2013.

\bibitem[\protect\citeauthoryear{Maas \bgroup \em et al.\egroup
  }{2013}]{maas2013rectifier}
Andrew~L Maas, Awni~Y Hannun, and Andrew~Y Ng.
\newblock Rectifier nonlinearities improve neural network acoustic models.
\newblock In {\em DLASLP at ICML}, 2013.

\bibitem[\protect\citeauthoryear{Mu \bgroup \em et al.\egroup
  }{2016}]{mu2016slink}
Xin Mu, Feida Zhu, Ee-Peng Lim, Jing Xiao, Jianzong Wang, and Zhi-Hua Zhou.
\newblock User identity linkage by latent user space modelling.
\newblock In {\em SIGKDD}, 2016.

\bibitem[\protect\citeauthoryear{Nassar \bgroup \em et al.\egroup
  }{2018}]{nassar18lowrank}
Huda Nassar, Nate Veldt, Shahin Mohammadi, Ananth Grama, and David~F. Gleich.
\newblock Low rank spectral network alignment.
\newblock WWW, 2018.

\bibitem[\protect\citeauthoryear{Ribeiro \bgroup \em et al.\egroup
  }{2017}]{ribeiro2017struc2vec}
Leonardo~F.R. Ribeiro, Pedro~H.P. Saverese, and Daniel~R. Figueiredo.
\newblock Struc2vec: Learning node representations from structural identity.
\newblock SIGKDD, 2017.

\bibitem[\protect\citeauthoryear{Singh \bgroup \em et al.\egroup
  }{2008}]{singh2008isorank}
Rohit Singh, Jinbo Xu, and Bonnie Berger.
\newblock Global alignment of multiple protein interaction networks with
  application to functional orthology detection.
\newblock {\em PNAS}, 2008.

\bibitem[\protect\citeauthoryear{Tan \bgroup \em et al.\egroup
  }{2014}]{tan2014elinkage}
Shulong Tan, Ziyu Guan, Deng Cai, Xuzhen Qin, Jiajun Bu, and Chun Chen.
\newblock Mapping users across networks by manifold alignment on hypergraph.
\newblock AAAI, 2014.

\bibitem[\protect\citeauthoryear{Wang \bgroup \em et al.\egroup
  }{2017}]{wang2017visiongan}
Jun Wang, Lantao Yu, Weinan Zhang, Yu~Gong, Yinghui Xu, Benyou Wang, Peng
  Zhang, and Dell Zhang.
\newblock Irgan: A minimax game for unifying generative and discriminative
  information retrieval models.
\newblock SIGIR, 2017.

\bibitem[\protect\citeauthoryear{Yu \bgroup \em et al.\egroup
  }{2017}]{yu2017nlpgan}
Lantao Yu, Weinan Zhang, Jun Wang, and Yong Yu.
\newblock Seqgan: Sequence generative adversarial nets with policy gradient.
\newblock AAAI, 2017.

\bibitem[\protect\citeauthoryear{Zhang and Philip}{2015}]{zhang2015multiple}
Jiawei Zhang and S~Yu Philip.
\newblock Multiple anonymized social networks alignment.
\newblock In {\em ICDM}, 2015.

\bibitem[\protect\citeauthoryear{Zhang and Tong}{2016}]{zhang2016final}
Si~Zhang and Hanghang Tong.
\newblock Final: Fast attributed network alignment.
\newblock In {\em SIGKDD}, 2016.

\bibitem[\protect\citeauthoryear{Zhou \bgroup \em et al.\egroup
  }{2016}]{zhou2016cycleconsistency}
Tinghui Zhou, Philipp Krahenbuhl, Mathieu Aubry, Qixing Huang, and Alexei~A
  Efros.
\newblock Learning dense correspondence via 3d-guided cycle consistency.
\newblock In {\em CVPR}, 2016.

\bibitem[\protect\citeauthoryear{Zhu \bgroup \em et al.\egroup
  }{2017}]{zhu2017cyclegan}
Jun-Yan Zhu, Taesung Park, Phillip Isola, and Alexei~A Efros.
\newblock Unpaired image-to-image translation using cycle-consistent
  adversarial networks.
\newblock In {\em ICCV}, 2017.

\end{thebibliography}

\end{document}